\documentclass{article}

\usepackage{PRIMEarxiv}

\usepackage[utf8]{inputenc} 
\usepackage[T1]{fontenc}    
\usepackage{hyperref}       
\usepackage{url}            
\usepackage{booktabs}       
\usepackage{amsfonts}       
\usepackage{nicefrac}       
\usepackage{microtype}      
\usepackage{lipsum}
\usepackage{fancyhdr}       
\usepackage{graphicx}       
\graphicspath{{media/}}     

\title{Navigating Complexity: Orchestrated Problem Solving with Multi-Agent LLMs
}

\author{
  Sumedh Rasal \\
  Georgia Institute of Technology \\
  Chicago, IL\\
  \texttt{srasal3@gatech.edu} \\
   \And
  E. J. Hauer \\
  \texttt{EJ@ThoughtCrafters.com} \\
}

\begin{document}
\maketitle

\begin{abstract}
Large Language Models (LLMs) have demonstrated remarkable capabilities in solving various tasks, yet they often struggle with comprehensively addressing complex and vague problems. Existing approaches, including multi-agent LLM systems, offer solutions to certain challenges but still require manual setup and lack scalability. To address this gap, we propose a novel approach leveraging decomposition to enable LLMs to tackle vague problems effectively.

Our approach involves an orchestrating LLM that interacts with users to understand the problem and then decomposes it into tangible sub-problems. Instead of expecting the LLM to solve the entire problem in one go, we train it to ask follow-up questions to gain a deeper understanding of the user's requirements. Once the problem is adequately understood, the orchestrating LLM divides it into smaller, manageable sub-problems. Each sub-problem is then assigned to specialized LLM agents or non-LLM functions for resolution. These agents work in parallel to solve their respective sub-problems, with the orchestrating LLM overseeing the process and compiling the solutions into a comprehensive answer for the user. By adopting this decomposition approach, we alleviate the constraints imposed by token limitations on LLM outputs and empower them to provide nuanced solutions to complex and ambiguous problems.

Through our approach, we aim to enable LLMs to think and operate more like humans, breaking down complex problems into manageable parts and collaboratively solving them. This not only enhances the problem-solving capabilities of LLMs but also offers a scalable and efficient method for addressing a wide range of real-world challenges.

\end{abstract}


\section{Introduction}
In recent years, the proliferation of Large Language Models (LLMs) has revolutionized problem-solving across various domains \cite{devlin2018bert} \cite{patil2023gorilla} \cite{touvron2023llama}. These models possess the ability to comprehend user queries and generate responses, often providing novel insights and solutions \cite{wei2022emergent}. Notably, LLMs exhibit exceptional performance in tasks that can be tackled through zero-shot or few-shot training techniques \cite{brown2020language}, such as transforming questions into familiar formats or solving coding problems based on analogous examples \cite{chen2021evaluating} \cite{sanh2021multitask}. As such, they hold immense potential for automating numerous tasks and streamlining workflows across diverse professional domains \cite{liu2023gpteval} \cite{chiang2023can} \cite{gao2023human} \cite{shen2023large} \cite{zheng2023judging}.

However, despite their remarkable capabilities, LLMs are not immune to limitations. Instances of hallucination, contradictory responses, or misinterpretation of user queries highlight the challenges inherent in relying solely on these models for problem-solving \cite{azamfirei2023large}. Existing frameworks designed to aid users in systematically addressing novel problems are constrained by token limits imposed by LLM models \cite{gur2023real}. These limits restrict the number of words or tokens that an LLM can process in a single message or query, often hindering the comprehensiveness of responses and impeding the continuity of conversations. While approaches such as chaining previous conversations or employing semantic search techniques mitigate some of these challenges, they are susceptible to errors, including redundancy or irrelevance in retrieved information \cite{ouyang2022training} \cite{saunders2022self} \cite{liu2023your}.

To address these limitations, a promising avenue lies in leveraging multi-agent frameworks to tackle lengthy or complex problems \cite{du2023improving} \cite{cohen2023lm} \cite{dafoe2021cooperative} \cite{dafoe2020open} \cite{kramar2022negotiation} \cite{yao2022react}. Such frameworks distribute problem-solving tasks among multiple LLM agents, offering a more robust approach compared to heuristic methods \cite{li2023camel} \cite{liang2023encouraging} \cite{qian2023communicative} \cite{lazaridou2020multiagent} \cite{graesser2020emergent} \cite{lee2018emergent} \cite{wu2023large}. However, the manual assignment of personas or roles to each LLM agent presents a bottleneck, particularly when dealing with novel problems lacking predefined roles \cite{rasal2024llm}.

In this paper, we propose a novel solution that integrates the strengths of existing approaches while addressing their limitations. Our approach centers on an orchestration strategy that employs problem decomposition to facilitate the seamless resolution of lengthy or complex problems. Inspired by human problem-solving approaches, our framework involves an orchestration LLM that engages in iterative conversations with users to ascertain problem requirements. Once the requirements are elucidated, the orchestration LLM decomposes the problem into manageable subproblems and assigns them to specialized agents, each tailored to address specific aspects of the problem. Notably, our approach incorporates non-LLM functionalities where necessary, enhancing robustness and adaptability to real-world tasks \cite{qin2023toolllm}. By leveraging problem decomposition and distributed task assignment, our framework offers a comprehensive and scalable solution for problem-solving with LLMs.

This is what we propose:
\begin{itemize}
\item Our proposed solution integrates problem decomposition and multi-agent orchestration, inspired by human problem-solving approaches, to address the limitations of existing frameworks.
\item By leveraging problem decomposition and distributed task assignment, our framework offers a comprehensive and scalable solution for problem-solving with LLMs, enhancing adaptability to real-world tasks.
\end{itemize}

\section{Methodology}

In response to these challenges, our methodology adopts a novel approach that seeks to address the token limit issue through algorithmic enhancements rather than relying solely on memory expansion. At the core of our methodology lies the development of an Orchestration LLM, tasked with the responsibility of comprehending user problem statements, decomposing them into manageable subproblems, and orchestrating interactions among specialized LLM agents.

Inspired by human problem-solving strategies, our methodology mirrors the process of breaking down complex tasks into smaller, more digestible components. We employ advanced natural language processing (NLP) techniques to analyze and dissect complex queries, identifying distinct subproblems across various domains. These subproblems are then mapped to specialized LLM agents, each pre-trained on specific tasks such as data analysis, creative writing, or technical problem-solving, to ensure precision and relevance in their responses.

Analogous to seeking expertise in real-life scenarios, our methodology leverages collaboration with domain experts to optimize problem-solving outcomes. For instance, just as individuals seek tax experts for efficient tax filing, our approach integrates specialized agents to tackle domain-specific challenges effectively.

Furthermore, we develop an algorithmic framework that enables the Orchestration LLM to decompose complex problems based on predefined criteria such as problem domain, required expertise, and task complexity. Finally, we implement a task assignment system to allocate subproblems among selected specialized agents, considering factors such as availability, expertise, and workload to streamline the problem-solving process and optimize resource utilization.

The high-level algorithm to our framework can be found \ref{tab:problem_solving}

\begin{tabular}{ll}
\toprule
\textbf{Step} & \textbf{Orchestration of Problem Solving with LLMs} \\
\midrule
1 & \textbf{Require} A complex problem statement \\
2 & \textbf{Ensure} A comprehensive solution to the problem \\
3 & \textbf{Begin} \\
4 & Input the complex problem statement \\
5 & Decompose the problem into subproblems \\
6 & \textbf{for} each subproblem \textbf{do} \\
7 & - Select a specialized LLM agent based on the subproblem's domain and requirements \\
8 & - Assign the subproblem to the selected agent \\
9 & - Agent solves the subproblem and stores the solution \\
10 & \textbf{end for} \\
11 & - Aggregate the solutions of subproblems into a coherent overall solution \\
12 & - Output the comprehensive solution \\
13 & \textbf{End} \\
\bottomrule
\label{tab:problem_solving}
\end{tabular}


\subsection{Decomposition Approach}
The implementation of our framework initiates with the user's submission of a complex problem statement, facilitated through a conversational interface within the LangChain framework \cite{chase2022langchain}. Within the Orchestration LLM, this input undergoes meticulous processing, wherein key information and requirements are extracted and analyzed with precision. Subsequently, the Orchestration LLM seamlessly applies a sophisticated decomposition algorithm, systematically breaking down the problem into manageable subproblems, thereby ensuring clarity and precision throughout the process.

The decomposition approach was crafted through a series of few-shot training iterations employing the GPT4 model \cite{openai2023gpt} \cite{bubeck2023sparks}. Initially, the LLM model decomposed the main problem into vague subproblems, which necessitated further refinement to ensure alignment with the expected decomposition format. Utilizing few-shot training, we provided the model with chain-of-thought type reasoning to guide its decomposition process effectively.

Upon successful decomposition, the Orchestration LLM assumes the pivotal role of assigning these subproblems to the most suitable specialized LLM agents. This assignment process is meticulously orchestrated to ensure that each subproblem is delegated to an agent possessing the requisite expertise and capabilities to effectively address it.

Central to the framework's efficacy is the decomposition phase, which serves as its cornerstone. We deliberately crafted the initial design to be adaptable and generalizable across diverse problem domains, prioritizing the demonstration of decomposition capabilities over specialized use-case system development.










\section{Experiments}
Our experiments center around the utilization of an orchestration LLM agent, employing OpenAI's gpt-4 model \cite{openai2023gpt}, which adeptly engages with users to elucidate requirements and subsequently decomposes problems into manageable subproblems. This orchestration LLM also determines the most suitable specialized LLM agent to address each subproblem. Each specialized LLM agent employed in our experiments is based on OpenAI's gpt3.5-turbo model, configured with a temperature parameter set to 0.0, reflecting the balance between coherence and creativity in responses. Facilitating communication among LLM agents is the LangChain framework, enabling seamless multi-agent interaction and facilitating focus on problem-solving.

\subsection{Experiment 1: Travel Planning with LLM Agents}
In this experiment, users present their travel plans to the framework, initiating a dialogue with the orchestration LLM to refine requirements and subsequently decompose the problem into subproblems. The orchestration LLM allocates tasks to specialized agents, which upon completion, contribute their findings for collation into a final course of action. Before execution, the orchestration LLM verifies details with the user to ensure accuracy.

Here's the conversational output of the framework.

\begin{quote}
\textbf{Initial Interaction: Gathering Requirements}

\textbf{User}: I need to book a return flight from Chicago Airport to Newark Airport from 3/22/24 to 3/26/24. I want to fly in luxury and travel between the hours of 10 am and 4 pm.

\textbf{Orchestration LLM Agent}: Got it. To ensure I find the best options for you, could you specify your preferences for seating class and any additional services you might require during your flight?

\textbf{User}: I prefer business class and would like access to in-flight Wi-Fi and a window seat.

\textbf{Orchestration LLM Agent Decomposes the Problem}

The Orchestration LLM Agent processes the user's request and decomposes it into subproblems to be handled by specialized agents:

\textbf{Flight Search Agent}: Find business class flights from Chicago to Newark that depart between 10 am and 4 pm on 3/22/24 and return flights on 3/26/24.

\textbf{Amenity Preferences Agent}: Identify airlines offering in-flight Wi-Fi and window seat availability in business class.

\textbf{Booking Agent}: Handle the booking process once the flight options are selected.

\textbf{Interaction with Specialized Agents}

\textbf{Orchestration LLM Agent to Flight Search Agent}: Find business class flights from Chicago to Newark departing between 10 am and 4 pm on 3/22/24 and returning on 3/26/24.

\textbf{Flight Search Agent}: Found several options.

\textbf{Orchestration LLM Agent to Amenity Preferences Agent}: Which of these options offers in-flight Wi-Fi and window seats in business class?

\textbf{Amenity Preferences Agent}: [Lists airlines that meet the criteria]
Finalizing the Option

\textbf{Orchestration LLM Agent to User}: Based on your preferences, here are the best flight options that offer business class, in-flight Wi-Fi, and window seats. [Presents options]

\textbf{User}: I'll take the option with Airline X on the 3/22 at 2 pm.

\textbf{Orchestration LLM Agent to Booking Agent}: Please book the flight with Airline X for 3/22, departing at 2 pm in business class, with a window seat and in-flight Wi-Fi.

\textbf{Booking Agent}: Your flight is booked. Here are your confirmation details.
\end{quote}

\subsection{Experiment 2: Exploratory Research and Summarization}
Users pose exploratory research queries to the framework, prompting the orchestration LLM to segment the problem into subproblems and assign tasks to specialized agents. Following the completion of tasks, the specialized agents provide their outputs, which are synthesized by the orchestration LLM into a comprehensive summary. Feedback from the user is sought to refine the final output as necessary.

Here's the conversational output of the framework.

\begin{quote}
\textbf{Initial Interaction: Gathering Requirements}

\textbf{User}: I need to research topics about long-term memory management in LLMs, focusing on what's currently possible and what's not. Then, I'd like a 5-page article summarizing these findings and suggesting future research directions.

\textbf{Orchestration LLM Agent Decomposes the Problem}

The Orchestration LLM Agent analyzes the request and identifies several key tasks:

\textbf{Literature Review Agent}: Conduct a comprehensive review of existing research on long-term memory management in LLMs.

\textbf{Analysis Agent}: Analyze the collected research to distinguish between solved problems, ongoing challenges, and unexplored areas.

\textbf{Writing Agent}: Synthesize the findings into a coherent 5-page article, highlighting current capabilities, and limitations, and suggesting future research directions.

\textbf{Interaction with Specialized Agents}

Task 1: Conduct Literature Review

\textbf{Orchestration LLM Agent to Literature Review Agent}: Identify and summarize key research articles on long-term memory management in LLMs, focusing on methodologies, findings, and limitations.

\textbf{Literature Review Agent}: [Returns a summary of articles, including key methodologies, findings, challenges, and limitations related to long-term memory management.]

Task 2: Analyze Research

\textbf{Orchestration LLM Agent to Analysis Agent}: Based on the literature review, analyze the current state of long-term memory management in LLMs, identifying what is possible, what remains challenging, and potential areas for future research.

\textbf{Analysis Agent}: [Provides an analysis distinguishing between the current capabilities and limitations of long-term memory management in LLMs, along with suggestions for future research directions.]

Task 3: Write Article

\textbf{Orchestration LLM Agent to Writing Agent}: Using the information collected and analyzed, write a 5-page article summarizing the findings on long-term memory management in LLMs. Highlight what is currently possible, and the limitations, and suggest ideas for future research.

\textbf{Writing Agent}: [Drafts a 5-page article that summarizes the state of long-term memory management in LLMs, including an overview of existing research, current limitations, and proposed future directions.]

Finalizing the Article

\textbf{Orchestration LLM Agent to User}: Based on the comprehensive research and analysis conducted by specialized agents, we have compiled a 5-page article summarizing the current state of long-term memory management in LLMs. The article discusses what is currently possible, outlines the main challenges and limitations, and suggests potential areas for future research.

\textbf{User}: Reviews the article and provides feedback or requests revisions if necessary.

\end{quote}

\section{Results}
Rather than relying solely on conventional datasets typically utilized in research, we opted to assess the efficacy of our framework using the GSM8K dataset, chosen to specifically demonstrate the capabilities of our approach. We employed four distinct types of agents in our experimentation: a single LLM agent powered by the GPT3.5-turbo model, two LLM agents supported by GPT3.5-turbo models, two LLM agents utilizing the GPT3.5-turbo model combined with the chain-of-thought and persona approach \cite{wei2022chain} \cite{rasal2024llm}, and finally, two LLM agents operating with the GPT4 model \cite{openai2023gpt} for orchestration and decomposition methodology and GPT3.5-turbo model for the specialized LLM agents, which involves breaking down the primary problem into subproblems. Our orchestration framework represents a notable advancement over existing methodologies, with detailed results presented in Table \ref{tab:table1}.

\begin{quote}
    "GSM8K consists of 8.5K high-quality grade school math problems created by human problem writers. We segmented these into 7.5K training problems and 1K test problems. These problems take between 2 and 8 steps to solve, and solutions primarily involve performing a sequence of elementary calculations using basic arithmetic operations (+ - / *) to reach the final answer. A bright middle school student should be able to solve every problem."\cite{cobbe2021gsm8k}
\end{quote}

Furthermore, our observations indicated a significant enhancement in accuracy when employing a calculator function as opposed to a math-specialized LLM agent. This improvement stems from the inherent limitations of LLMs in executing mathematical operations, despite their versatility across various domains. We contend that rather than training LLMs for math operations, our framework facilitates the orchestration LLM in discerning whether a subproblem necessitates an LLM agent or an alternative non-LLM approach.

\begin{table}
 \caption{GSM8K Data set Evaluation}
  \centering
  \begin{tabular}{lll}
    \toprule
    Agent     & Solve Rate     \\
    \midrule
    Single GPT3.5-turbo & 50\% \\
    Multi-Agent GPT3.5-turbo & 55\% \\
    Multi-Agent GPT3.5-turbo (LLM-Harmony) & 65\% \\
    Multi-Agent GPT4 \textbf{(Our Approach)} & \textbf{73\%} \\
    \bottomrule
  \end{tabular}
  \label{tab:table1}
\end{table}

\section{Limitations}
While our decomposition framework represents a significant advancement, it possesses inherent limitations that warrant consideration. Foremost among these limitations is the framework's lack of generalization to address all conceivable problem scenarios, both novel and otherwise. We recognize the necessity of further training the orchestration model, essentially serving as a foundational Large Language Model (LLM), to acquire a comprehensive understanding of diverse problem-solving approaches. Presently, the model exhibits proficiency in decomposing various problems, yet its scope may not extend to accommodate the complexities of future challenges.

For instance, consider a user's request for transportation. While the current model adeptly decomposes the problem into a specialized agent capable of booking a ride, scenarios may arise where multiple specialized agents possess this capability. In such instances, the orchestration LLM faces the challenge of determining the most appropriate course of action. Additionally, the framework lacks a mechanism to assimilate information about new specialized agents, highlighting the need for a retraining mechanism to continually update the orchestration LLM's knowledge base.

Furthermore, while the current model effectively decomposes problems, it lacks the capability to iteratively learn and refine its problem-solving strategies based on the outcomes of decomposition steps. Addressing this limitation represents a key focus of our future research endeavors. Nonetheless, the primary objective of this paper is to disseminate our decomposition approach to the readership, offering insights into its application in problem-solving contexts.

\section{Conclusion}
In conclusion, our paper introduces a novel approach to problem-solving using a decomposition framework orchestrated by Large Language Models (LLMs). Through meticulous decomposition of complex problem statements into manageable subproblems, our framework demonstrates the potential to streamline problem-solving processes and mitigate the limitations inherent in current LLM-based solutions. By leveraging the inherent capabilities of LLMs to understand user queries and decompose problems, our approach offers a promising avenue for addressing a wide range of challenges across various domains.

Our implementation of the framework highlights its efficacy in breaking down intricate problem statements and assigning tasks to specialized LLM agents, resulting in coherent and comprehensive solutions. Moreover, our experiments underscore the versatility of the framework in facilitating tasks such as travel planning and research summary generation, showcasing its applicability in real-world scenarios.

However, our work also reveals several limitations that necessitate further research and development. While the decomposition framework represents a significant advancement, its generalization to novel problem scenarios remains an ongoing challenge. Additionally, the framework lacks mechanisms for adaptive learning and integration of new specialized agents, indicating avenues for future improvement.

Despite these limitations, our paper contributes to the growing body of literature on LLM-based problem-solving approaches by introducing a systematic and scalable framework. By sharing our approach and insights, we aim to stimulate further research and innovation in this domain, paving the way for enhanced problem-solving capabilities and the realization of artificial general intelligence.


\bibliographystyle{apalike}
\bibliography{references}  

\end{document}